\documentstyle[12pt]{article}

\newcommand{\scsi}{\scriptsize}

\newcommand{\etal}{\mbox{\it et al.\/}}

\newcommand{\ssn}{$^{1}S_{0}$}

\newcommand{\tpt}{$^{3}P_{2}$}

\newcommand{\tft}{$^{3}F_{2}$}
\newcommand{\sdt}{$^{1}D_{2}$}

\newcommand{\tfd}{$^{3}F_{3}$}

\newcommand{\Tlab}{\mbox{$T_{\mbox{\scsi lab}}$}}

\newcommand{\chis}{\mbox{$\chi^{2}$}\/}

\newcommand{\chismin}{\mbox{$\chi^{2}_{\mbox{\scsi min}}$}}

\newcommand{\Ndf}{\mbox{$N_{\mbox{\scsi df}}$}}

\newcommand{\VS}{\mbox{$V_{\mbox{\scsi S}}$}}
\newcommand{\VL}{\mbox{$V_{\mbox{\scsi L}}$}}
\newcommand{\VR}{\mbox{$V_{\mbox{\scsi R}}$}}
\newcommand{\VI}{\mbox{$V_{\mbox{\scsi I}}$}}

\newcommand{\VEM}{\mbox{$V_{\mbox{\scsi EM}}$}}
\newcommand{\VNUC}{\mbox{$V_{\mbox{\scsi NUC}}$}}

\begin{document}
\title{Partial Wave Analyses of the $pp$ data alone \\
       and of the $np$ data alone}
\author{R.A.M.M. Klomp, J.-L. de Kok, M.C.M. Rentmeester, \\
        Th.A. Rijken, and J.J. de Swart \\
        Institute for Theoretical Physics \\
        Nijmegen, The Netherlands}
\date{}
\maketitle
\begin{abstract}
We present results of the Nijmegen partial-wave analyses of all $N\!N$
scattering data below $\Tlab = 500$ MeV. We have been able to extract
for the first time the important $np$ phase shifts for both $I = 0$ and
$I = 1$ from the $np$ scattering data alone. This allows us to study the
charge independence breaking between the $pp$ and $np$ $I = 1$ phases.
In our analyses we obtain for the $pp$ data $\chismin/\Ndf = 1.13$ and for
the $np$ data $\chismin/\Ndf = 1.12$.
\end{abstract}
\section*{Introduction}
The last 15 years the Nijmegen group has been working on partial-wave
analyses (PWA) of the $N\!N$-scattering data for energies below
$\Tlab = 350$ MeV. In these analyses, the $pp$ data furnish the $I = 1$
phase shifts~\cite{Ber90} in the lower partial waves with $J \leq 4$.
It has been customary to do these low energy PWA up to $\Tlab = 350$ MeV,
despite of the fact that the inelasticities set in at $\Tlab \simeq 280$ MeV.
It can be shown, however, that up to $\Tlab = 350$ MeV the inclusion of
inelasticities in $pp$ scattering improves the total $\chis \simeq 1787$
with only about 1, so neglecting the inelasticities is totally warranted.

In the PWA of all the $np$ data below $\Tlab = 350$ MeV, it has always
been impossible to determine all important phase shifts, when only the $np$
data were used. The standard procedure has therefore been to take the
$I = 1$ phases (except the \ssn) over from the $pp$ analyses, with or
without corrections for Coulomb, and $np$ and $\pi^0\pi^\pm$ mass
difference effects. The \ssn\ $np$ phase shift was always searched for.
It was found that there is a definite charge independence breaking in these
\ssn\ phases. An attempt by us, to extract all the lower $I = 0$ and
$I = 1$ phase shifts in an analysis of all $np$ data below $\Tlab = 350$ MeV
failed. However, it was possible to determine the $np$ $^3P$ waves, when
the higher partial waves were taken over from the $pp$ analysis.

Recently, the Nijmegen group extended the $pp$ PWA to energies so far above
the pion production thresholds, that the inclusion of inelasticities was
necessary. A preliminary version of such a PWA for all $pp$ data with
energies below $\Tlab = 500$ MeV has already been presented~\cite{Kok93}.

When the $np$ PWA was extended to energies up to $\Tlab = 500$ MeV, it
turned out to be possible to determine uniquely all the important partial
waves. Comparing then with the analogous $pp$ analysis gives indications
for possible charge independence breaking effects in other waves besides
the \ssn\ waves.

\section*{The method of analysis}
In the $np$ as well as in the $pp$ partial wave analyses, the various
phases are obtained by solving the relativistic Schr\"odinger equation
with a well-known long-range potential $\VL = \VNUC + \VEM$ for
$r \geq b = 1.4$ fm. This \VL\ contains the electromagnetic interaction
(such as Coulomb, magnetic moment, and vacuum polarization), the OPEP
tail, and the rest of the long range $N\!N$ interaction as given by the
Nijmegen potential~\cite{Nag78}.

For $r < b$ the short-range interaction is described by an energy-dependent,
square-well potential
\[
\VS = \VR - i\VI \ .
\]
For \VR\ we follow the same procedure as in the Nijmegen PWA and write
\[
\VR = \sum_{n=0}^{N} a_n (k^2)^n \ .
\]
For the imaginary part of the potential we take
\[
\VI = (k^2 - k_{\mbox{\scsi thr}}^2)^n V \cdot \theta(E - E_{\mbox{\scsi thr}})
 \ .
\]
\section*{Results}
\begin{figure}
\vspace{9cm}
\includegraphics{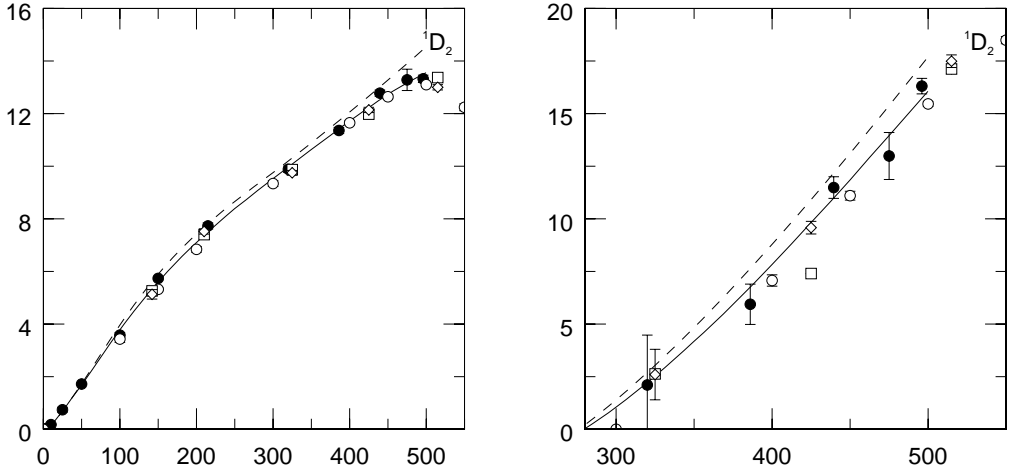}
\caption{Preliminary multienergy phase shifts and inelasticities in
the \protect\sdt\ partial wave in degrees vs. $\protect\Tlab$ in MeV.
Solid line: 0-500 MeV $pp$ partial-wave analysis;
dashed line: 0-500 MeV $np$ partial-wave analysis.
$\protect\bullet$: $pp$ single-energy analyses;
$\protect\circ$: Arndt \protect\etal~\protect\cite{Arn92};
$\protect\Box$: Dubois \protect\etal~\protect\cite{Dub82};
$\protect\diamond$: Bugg \protect\etal~\protect\cite{Bug90}.}
\label{dwave}
\end{figure}

The results of our PWA can be summarized as follows. The phase parameters
were parametrized with 36 parameters in the $pp$ analysis and 38 parameters
in the $np$ analysis. We allowed for up to four parameters in the real part
of the potential in each partial wave, which was found to be enough.
The actual number of parameters per partial wave varies from
four in the \ssn\ to one or none in the higher $G$ waves.
An imaginary part of the potential was only used in the \sdt\, and \tfd\
partial waves and in the $pp$ PWA also in the coupled \tpt-\tft\ partial waves.
As an example, Fig.~\ref{dwave} shows preliminary results for the \sdt\
partial wave.
We reach $\chismin = 3555.4$ for 3455 $pp$ scattering data and
$\chismin = 4142.0$ for 3968 $np$ scattering data.

\end{document}